## Title
- Anti-Sisyphus driving in a matter-wave swing.

## Authors

Wen L. Liu,[1,2] Jun Jian.,[1] Ning X. Zheng,[1] Hui Tang.,[4] Ji Z. Wu,[1,2], Yu Q. Li,[1,2] Wen X. Zhang,[3]* Jie Ma,[1]* and Suo T. Jia[1,2]

## Affiliations

1, *State Key Laboratory of Quantum Optics Technologies and Devices, Institute of Laser Spectroscopy, College of Physics and Electronics Engineering, Shanxi University, 030006, China.*
2, *Collaborative Innovation Center of Extreme Optics, Shanxi University, Taiyuan 030006, China*
3, *School of Physics, Hangzhou Normal University, Hangzhou, Zhejiang 311121, China*
4, *Shenzhen Institute for Quantum Science and Engineering, Southern University of Science and Technology, Shenzhen, Guangdong 518055, China*

*\* Corresponding author. E-mail: wxzhang@hznu.cn; mj@sxu.edu.cn;*

## Abstract

- Dilute-gas Bose-Einstein condensates are an exceptionally versatile testbed for the investigation of physics phenomenon especially the well-known classical system. Here we use a degenerate Bose gas of sodium atoms confined in an optical dipole trap to simulate the matter-wave on the swing. Under the driving of Anti-Sisyphus process, the swing was excited successfully. Moreover, the spin echo like behavior and collective-mode excitation appear during the oscillation of matter-wave swing, manifesting the quantum nature of the system beyond its classical counterpart. Our work lays the foundation for matter-wave on the swing and more generally points to a future of practical applications for the motional quantum states

    linked with quantum information science.

## MAIN TEXT
## Introduction

Quantum simulators play a promising role on the spectrum of quantum devices from specialized quantum experiments to universal quantum computers (1). These quantum devices utilize entanglement and many-particle behaviors to explore and solve hard scientific with inaccessible system or in extreme experimental condition, engineering, and computational problems (2). Such devices like defect-free arrays from initially stochastically loaded optical traps, recently developed by atom-by-atom assembly technique (3-5), has led to the most recent Rydberg quantum simulation applications (6).

In addition, Quantum simulators can also give the instructiveness of understanding the relation and boundary between quantum and classical systems (7–9). The quantum gas confined in the optical trap is one of the most important quantum simulators in modern condensed matter physics (10).

The quantum gas has been used to research the superfluid and supersolid (11), the matter-wave solitons (12), and other well-known physics phenomenon (13–15) or classical system such as Newton's cradle (16). Generalization of the Newton's cradle to quantum mechanical particles makes it easily to understand the time evolution of non-equilibrium trapped 1D Bose gases (17). Besides that, the separation of time scales in a Quantum Newton's Cradle was theoretically proposed (18).

Another classical system in the daily life is the swing. To start a classical swing, one needs to shift it from its equilibrium position or push it initially. To further increase its oscillation amplitude, one must keep driving it, for instance, pushing the swing periodically, with external forces. As a comparison, a quantum "swing", in a vacuum state, may be started from its ground state through parametric resonance, which deepens our understanding of the principle of quantum uncertainty (9). As to increasing the swing's oscillation amplitude, we propose to utilize anti-Sisyphus driving, which does not require any external force at all.

In this work, we study the Anti-Sisyphus driven matter-wave swing in a spinor Bose–Einstein condensate (BEC). The anti-Sisyphus process use the external magnetic field and a series of rf pulses. It is on the other direction of Sisyphus cooling that was extended to cool the alkaline earth atom in a tweezer successfully (4, 21). The condensate is made from sodium atoms carrying a hyperfine spin F = 1 confined in a spin independent optical dipole trap. The BEC stays in the dipole trap initially after force evaporation and then excited by the Anti-Sisyphus process.

## Results
### Experimental Setup

The experimental setup of the matter-wave on the swing is sketched in Fig. 2(a). We start our experiments with a Bose–Einstein condensate (BEC) on the F=1 state loaded into a red-detuned crossed dipole trap, with total atom number around $N = 10^5$. Two crossed dipole trap beams were all in the horizontal direction with $45^0$.

The beam waists were 33μm and 39μm, measured by the parametric heating method under high (14W) dipole trap laser power, respectively. The gradient magnetic field along the gravity direction kick the atoms and separate the spin components. A BEC of 5∗104 sodium atoms fully polarized into the -1 state in a crossed optical trap was achieved with the strong magnetic field, which kicks the atoms on the $m_F = 0, 1$ state out of the optical trap, leaving all atoms on the $m_F = -1$ state (22).

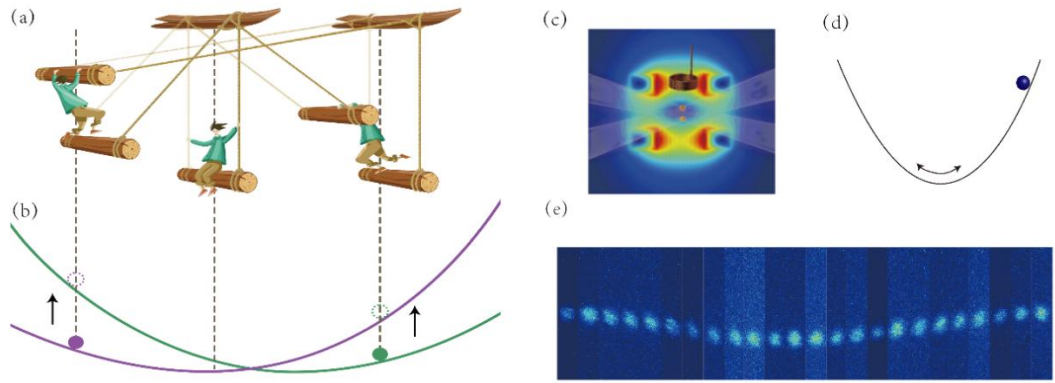

**Fig.1. (a)** Diagram of two classical swings. **(b)** Sketches at various times of clouds of atoms in two harmonic traps shifted by magnetic field. The two potentials are shifted by the magnetic field. The spin -1 state atoms are in the blue potential and transferred to +1 state by a π pulse. The atom cloud in +1 state oscillate in the purple trap after the 1$^{st}$ π pulse. The second pulse is induced when the atom cloud climbs to the maximum position. The atoms are transferred to the -1 state again. The movement occurs again in the blue potential. **(c)** Experimental setup. A Bose–Einstein condensate of spin-1 sodium atoms was held in a crossed dipole trap. Magnetic field produced by magnetic coils separate the -1 state and +1 state atom into two potentials. **(d)** The BEC show a damped swing movement type in the potential formed by one crossed optical dipole trap. **(e)** Selected images showing the free evolution of +1 state atom cloud in the potential from one side to another side.

We measure the linear integrated densities mF in the trap after a short expansion in a magnetic field gradient that separates all three components $m_F = 0, \pm 1$ by the Stern–Gerlach effect. In order to measure the density distributions, absorption images were acquired for the condensates after undergoing free expansion for 8ms. The −1 state atoms are in the blue potential shown in the Fig.1(b), and almost in their minimum position. A π pulse is induced and transfer the atoms to $m_F = +1$ state. Under the action of a big offset magnetic field, the potentials for BEC in the +1 and −1 state are shifted separately, as the purple and blue line shown in the Fig.1(b). In the $m_F =+1$ hyperfine state, the movement of atom cloud can only occur in the purple potential, which means that the atoms are on the non-equilibrium position of the purple trap. After a quarter oscillation period, the atoms will be at the bottom of the swing, but its momentum will be maximal. The matter-wave on the swing was excited for the first time. As the decay oscillation schematic shown in the Figure 2(b), the +1 state atoms cloud climb to another non-equilibrium position, where the momentum of the atoms is zero.

We introduce another new π pulse, the second π pulse. The atoms were transferred to −1 state by this pulse again and start to slide down in the blue trap. The selected images in the Fig.2(c) show the free evolution of +1 state atoms in the potential. Different from the process reported by the Ref. (23), the amplitude of oscillation increase means the atom cloud climb to a new higher position in the potential. Such a new position makes the atom cloud feel a bigger magnetic field, as the Zeeman splitting between magnetic sublevels was increased, the frequency for next π pulse had to be enlarged. We stop the driving process by seven pulses. The damping oscillations of the +1-state atom cloud excited by the pulses are shown in the Fig.2.

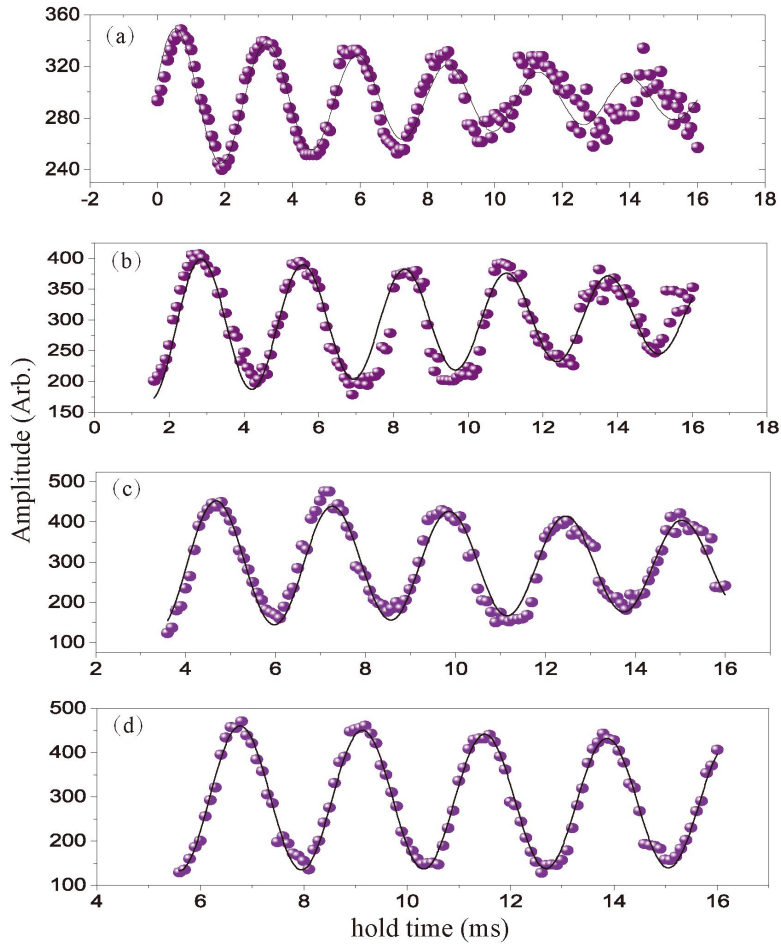

**Fig. 2**. **(a-d)** Dependence of the free oscillation of atom in +1 state in the potential on the number of RF pulses. The pulses number are 1, 3, 5, 7 for the Fig.2(a-d), separately. The amplitude increases significantly because of the Anti-Sisyphus driven. Each swing shows damped oscillation during the free evolution. The red lines are the damped sine fitting results.

**Analysis**

The free oscillations after all pulses in the Fig.2 can be described by the exponentially damped sinusoidal function superposed onto a linear decrease:

$$Y = A e^{-t/\tau} \sin(2\pi \omega t + \Phi) + ct + q$$

Here, $A$ is amplitude of oscillation, $\tau$ is damping constant, t is free oscillation time, $\omega$ is oscillation frequency, $\phi$ is initial phase, q is the average offset. The matter-wave swing was excited higher and higher after each pulse. The amplitudes vs number of pulses are shown in the Fig.3(a). The amplitude for +1 state and −1 states are shown together. The red line is linear fitting. The linear increase of its amplitude of the matter wave swing is similar with the classical swing shown in Fig.1(a). The oscillation amplitude reduces during the free evolution after all pulses, due to the damping of the environment. However, combined with the action of anti-Sisyphus driving process, the swing shows slower amplitude damping with the number of $\pi$ pulses.

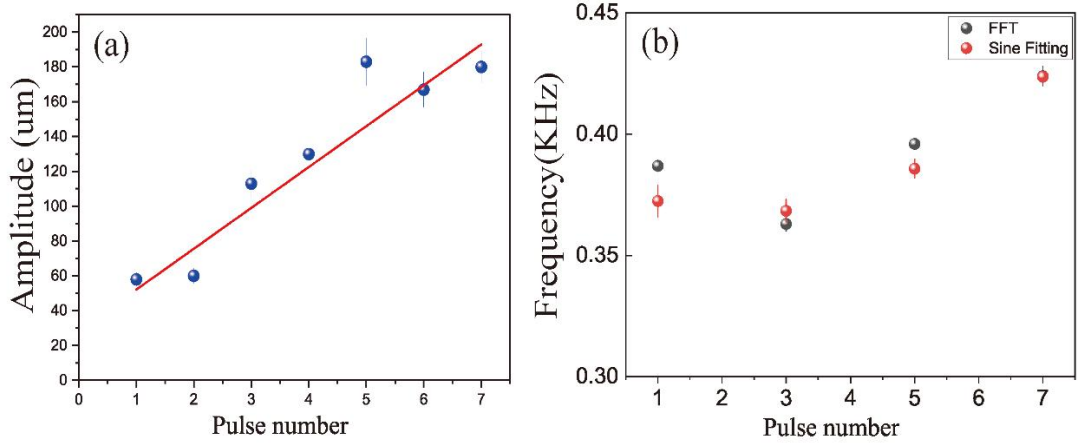

**Fig. 3**. **(a)** The amplitude for the oscillations after action of each pulse, red line is the linear fitting. **(b)** The oscillation frequency of Fig.3(a-d) analysis by damped sinusoidal fitting and FFT.

We thus fit the oscillations with two functions, the exponentially damped sinusoidal function and (Fast Fourier Transform) FFT. The fitting frequency for the free evolution of atom cloud in a potential was shown in the Fig.3(b), the frequency is around $v_{axial}$ = 390 Hz. As the Fig.3(b) shown, the red point is the result fitted by damped sinusoidal function, the black results are got by FFT fitting. Comparing with the two analysis results, the maximum difference is about 3.7 Hz. The anti-Sisyphus driving provides a possible cross-check method to measure the trap frequency of tightly focused trap, especially for the optical tweezers (24).

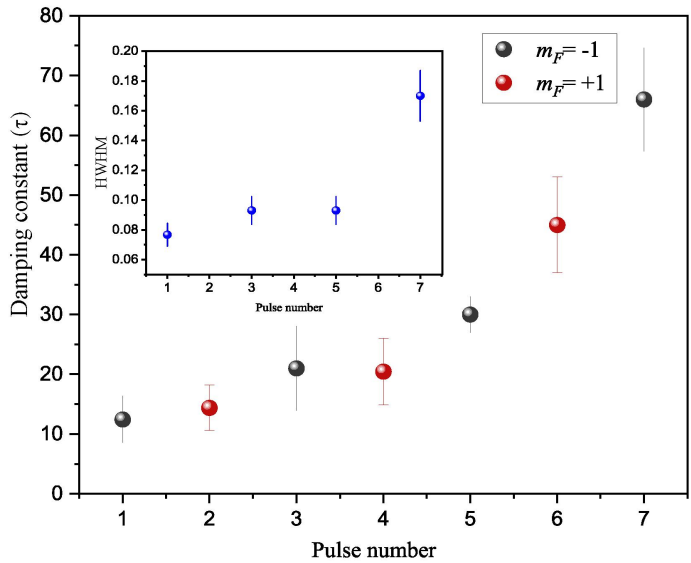

**Fig.4**. Damping constant by the damped sinusoidal fitting. The slowly decay means that the spin echo like effect is induced during the Anti-Sisyphus driving. Insert is the HWHM of FFT fitting for the data in the Fig.2(a-d).

Considering the damping free evolution shown in the Fig2. Damping constant $\tau$ for each pulses shown in the Fig.4 are got from the Equation 1. The constant increases with the pulses mean that the decay for the oscillation goes slowly, which means the spin echo like phenomenon is induced in the swing. The decay constants got by analyzed with FFT are also shown by the insert figure in the Fig.4, which have the similar behavior with the results of Fig.4. The spin echo was induced after several pulses. The results by analyzing both with the FFT and damped sin fitting show that the spin echo can be increased by multi pulses. In the experiment, limited by the stability and precision of the magnetic field, the atoms were lost and heated during the driving process, the number for π pulse is set to be 7. There is a possible to apply more pulses while less heating and loss in order to reduce the decay with more than 7 pulses. The spin echo effect can be enhanced more clearly by multi-pulses. The spin-echo reverses the inhomogeneous dephasing of a spin ensemble, which significantly increases the coherence time of the quantum system.

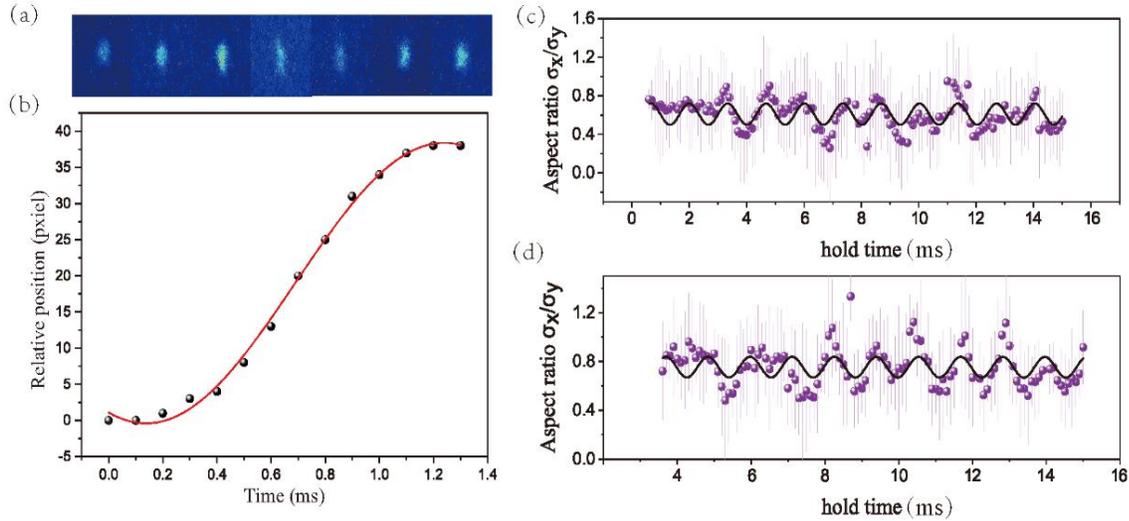

**Fig. 5. (a)** Shape oscillation of a Bose-Einstein condensate. Absorption images of condensates at several free oscillation times are shown as examples. The absorption images were taken after the switch-off of the trapping potential with 8ms of ballistic expansion. **(b)** The corresponding relative positions of the atom cloud in the potential. **(c, d)** Variation of the aspect ratio of the condensate in +1 (−1) state, excited by the third (sixth) π pulse separately, plotted as a function of the free oscillation time in the trap. The average frequency of the aspect ratio oscillations obtained from the sinusoidal fit is around 770Hz, for a cigar-shape BEC.

The condensate was allowed to evolve in the potential for various free oscillation. The atom cloud shows significantly squeezed during the free evolution in the trap under the action of third pulse as shown in the Fig5(a). Fig.5(b) is the corresponding positions of the atoms in the trap with 0.2ms interval. The red line is the damped sinusoidal fitting. Collective-mode excitation of the dilute Bose condensate, shown in the absorption images in Fig.5(c-d), is observed from the matter wave on the swing. Here, the shape oscillation was in moment space, as we observe the velocity of the condensate, and the vertical (horizontal) width reflects the expansion. We analyzed the time evolution of the vertical and horizontal widths of the condensates to estimate the frequency of collective oscillation in detail through a series of time-fly images. The widths were obtained by fitting the density profiles to parabolic function. The aspect ratio $\sigma_x/\sigma_y$ in the trap for the +1(−1) state were shown in the Fig.5 (c-d) respectively. We only take the ratio instead of the $\sigma_x$ or $\sigma_y$, as the atom number is not so stable. We plot the aspect ratio as a function of the free oscillation time. The data were fitted to an exponential sine function, giving an oscillation frequency of $2\nu_{radial}$, in good agreement with the theoretical prediction (25). The collective excitation mode through the Anti-Sisyphus driven matter wave swing was shown in this paper clearly. Such collective excitations may be caused by the anharmonic property of the trap edge. The atom cloud moves to the edge of the trap when pulse was induced, the confine is not the same.

**Discussion**

We drive the matter wave on the swing with Anti-Sisyphus process. The spin echo like phenomenon and collective-mode excitation appear during the process. We had tried to excite the atom cloud with higher energy, in order to observe the transition from collective to single particle behavior as the Ref. (25) had reported, while, when the pulse number arrive at eight, the atomic cloud is almost broken. It is hard to drive the swing with more than seven pulses.

Being inspired by an atom swing, reported by the Ref (20), with the rapid development of single atom tweezer manipulation, we can drive the matter wave with an atom instead of many atoms to reconstruct the motional quantum states. In addition, the matter-wave swing can be used as a start to explore squeezing experiments (26, 27) and an interesting method to do a high precision measurement (28).

**Materials and Methods**

The experimental details to get spinor BEC can be found from Ref (29,30). Hot atoms are slowed down by a Zeeman slower and then collected in a magneto-optical trap (MOT). After compassed MOT (cMOT) and Molasses process, $6 \times 10^8$ atoms are cooling to ~35μk. The atoms are then loaded in to a crossed red detuning optical dipole trap (ODT) to be forced evaporation. A pure F = 1 BEC of $1.6 \times 10^5$ sodium atoms is created from a 3.5s force evaporation (30). To control the initial spin population, we apply different magnetic field gradient during the evaporation. To create a pure F=1, mF = 0 condensate, a field of 10 G/cm is applied during the final evaporation (22,31).

In the Anti-Sisyphus driven process, the atoms climb to higher position, the Zeeman energy shifts are enlarged by the magnetic field. The frequency of these RF pulses is not the same. In order to set the right RF pulse, the Rabi oscillation for each pulse is measured. All the experimental data are fitted by the bimodal (Gaussian + parabola) distribution, both the position evolution information and the atom cloud size are got from the fitting.

**References**
1. P. Scholl, M. Schuler, H. J. Williams, A. A. Eberharter, D. Barredo, N. Schymik, Kai, V. Lienhard, L.-P. Henry, T. C. Lang, T. Lahaye, A. M. Läuchli, A. Browaeys, Quantum simulation of 2d antiferromagnets with hundreds of Rydberg atoms, Nature **595**, 1476 (2021).
2. R. Finkelstein, R. B.-S. Tsai, X. Sun, P. Scholl, S. Direkci, T. Gefen, J. Choi, A. L. Shaw, and M. Endres, Universal quantum operations and ancilla-based read-out for tweezer clocks, Nature **634**, 321 (2024).
3. A. L. Shaw, P. Scholl, R. Finklestein, I. S. Madjarov, B. Grinkemeyer, M. Endres, Dark-state enhanced loading of an optical tweezer array, Phys. Rev. Lett. **130**, 193402 (2023).
4. A. Cooper, J. P. Covey, I. S. Madjarov, S. G. Porsev, M. S. Safronova, M. Endres, Alkaline-earth atoms in optical tweezers, Phys. Rev. X **8**, 041055 (2018).
5. S. Saskin, J. T. Wilson, B. Grinkemeyer, J. D. Thompson, Narrow-line cooling and imaging of ytterbium atoms in an optical tweezer array, Phys. Rev. Lett. **122**, 143002 (2019).
6. T. Browaeys, A. Lahaye, Many-body physics with individually controlled Rydberg atoms, Nat. Phys. **16**, 1745 (2020).

7. X. Chai, D. Lao, K. Fujimoto, R. Hamazaki, M. Ueda, C. Raman, Magnetic solitons in a spin-1 bose einstein condensate, Phys. Rev. Lett. **125**, 030402 (2020).
8. T. M. Bersano, V. Gokhroo, M. A. Khamehchi, J. D'Ambroise, D. J. Frantzeskakis, P. Engels, P. G. Kevrekidis, Three-component soliton states in spinor f = 1 bose-einstein condensates, Phys. Rev. Lett. **120**, 063202 (2018).
9. C. M. Wilson, G. Johansson, A. Pourkabirian, M. Simoen, J. R. Johansson, T. Duty, F. Nori, P. Delsing, Observation of the dynamical casimir effect in a superconducting circuit, Nature **479**, 376–379(2011).
10. J. Simon, W. S. Bakr, R. Ma, M. E. Tai, P. M. Preiss, M. Greiner, Quantum simulation of antiferromagnetic spin chains in an optical lattice, Nature **472**, 1476 (2011).
11. M. Boninsegni, N. V. Prokof è v, Colloquium: Supersolids: What and where are they ?, Rev. Mod. Phys. **84**,759 (2012).
12. J. H. V. Nguyen, P. Dyke, D. Luo, B. A. Malomed, R. G. Hulet, Collisions of matter-wave solitons, Nat. Phys. **10**, 1745 (2014).
13. A. Farolfi, D. Trypogeorgos, C. Mordini, G. Lamporesi, G. Ferrari, Observation of magnetic solitons in two component bose-einstein condensates, Phys. Rev. Lett.**125**, 030401 (2020).
14. S.-L. Zhu, H. Fu, C.-J. Wu, S.-C. Zhang, L.-M. Duan, Spin hall effects for cold atoms in a light-induced gauge potential, Phys. Rev. Lett. **97**, 240401 (2006).
15. R. Sajjad, J. L. Tanlimco, H. Mas, A. Cao, E. Nolasco-Martinez, E. Q. Simmons, F. L. N. Santos, P. Vignolo, T. Macrì, D. M. Weld, Observation of the quantum boomerang effect, Phys. Rev. X **12**, 011035 (2022).
16. R. Sajjad, J. L. Tanlimco, H. Mas, A. Cao, E. Nolasco-Martinez, E. Q. Simmons, F. L. N. Santos, P. Vignolo, T. Macrì, and D. M. Weld, A quantum newton's cradle, Nature **440**, 1476 (2006).
17. Y. Tang, W. Kao, K.-Y. Li, S. Seo, K. Mallayya, M. Rigol, S. Gopalakrishnan, and B. L. Lev, Thermalization near integrability in a dipolar quantum newton's cradle, Phys. Rev. X **8**, 021030 (2018).
18. R. van den Berg, B. Wouters, S. Eliëns, J. De Nardis, R. M. Konik, J.-S. Caux, Separation of time scales in a quantum newton's cradle, Phys. Rev. Lett. **116**, 225302 (2016).
19. M. Brown, S. Muleady, W. Dworschack, R. Lewis-Swan, A. Rey, O. Romero-Isart, C. Regal, Time-of-flight quantum tomography of an atom in an optical tweezer, Nat. Phys. **19**, 569 (2023).
20. H. Bernien, A picture of a swinging atom, Nat. Phys. **19**, 474–475 (2023).
21. J. P. Covey, I. S. Madjarov, A. Cooper, M. Endres, 2000-times repeated imaging of strontium atoms in clock-magic tweezer arrays, Phys. Rev. Lett. **122**, 173201 (2019).
22. C. Käfer, R. Bourouis, J. Eurisch, A. Tripathi, H. Helm, Ejection of magnetic-field-sensitive atoms from an optical dipole trap, Phys. Rev. A **80**, 023409 (2009).
23. Y. Eto, M. Sadgrove, S. Hasegawa, H. Saito, T. Hirano, Control of spin current in a bose gas by periodic application of $\pi$ pulses, Phys. Rev. A **90**, 013626 (2014).
24. D. Ohlde Mello, D. Schäffner, J. Werkmann, T. Preuschoff, L. Kohfahl, M. Schlosser, G. Birkl, Defect-free assembly of 2d clusters of more than 100 single-atom quantum systems, Phys. Rev. Lett. **122**, 203601 (2019).
25. M.-O. Mewes, M. R. Andrews, N. J. van Druten, D. M. Kurn, D. S. Durfee, C. G. Townsend, W. Ketterle, Collective excitations of a bose-einstein condensate in a magnetic trap, Phys. Rev. Lett. **77**, 988 (1996).
26. F. Anders, A. Idel, P. Feldmann, D. Bondarenko, S. Loriani, K. Lange, J. Peise, M. Gersemann, B. Meyer-Hoppe, S. Abend, N. Gaaloul, C. Schubert, D. Schlippert, L. Santos, E.

Rasel, C. Klempt, Momentum entanglement for atom interferometry, Phys. Rev. Lett. **127**, 140402 (2021).
27. Y. Wu, R. Krishnakumar, J. Martínez-Rinćon, B. K. Malia, O. Hosten, M. A. Kasevich, Retrieval of cavity-generated atomic spin squeezing after free-space release, Phys. Rev. A **102**, 012224 (2020).
28. S. S. Szigeti, S. P. Nolan, J. D. Close, S. A. Haine, High-precision quantum-enhanced gravimetry with a bose-einstein condensate, Phys. Rev. Lett. **125**, 100402 (2020).
29. J. Jiang, L. Zhao, M. Webb, N. Jiang, H. Yang, Y. Liu, Simple and efficient all-optical production of spinor condensates, Phys. Rev. A **88**, 033620 (2013)
30. Wen L. Liu, Ning X. Z., Xiao. F. W., Jing X., Yu Q. L., V. B. Sovkov, P. Li, Y. M. Fu, Ji Z. Wu, Jie M., Lian T. X., Suo T. Jia, Fast, simple, all-optical production of sodium spinor condensates, J. Phys. B: At. Mol. Opt. Phys. **54** 155501 (2021)
31. M.-S. Chang, C. D. Hamley, M. D. Barrett, J. A. Sauer, K. M. Fortier, W. Zhang, L. You, M. S. Chapman, Observation of Spinor Dynamics in Optically Trapped 87Rb Bose-Einstein Condensates, Phys. Rev. Lett. **92**.140403 (2004)

**Funding:**
This work is supported by the Innovation Program for Quantum Science and Technology (Grant No.2021ZD0302103), the National Natural Science Foundation of China (Grant Nos.62325505, 62020106014, 62175140, 62475138,62422508, 12274331), the National Key Research and Development Program of China (2022YFA1404201), and Graduate education innovation project of Shanxi Province (2024KY019).

**Author contributions:**
J. M, W. X. Z., and S. T. J. conceived and supervised the experiment. W. L. L., J. J., and N. X. Z. conducted the measurements. H. T., J. Z. W. and W. X. Z analyzed the data. W. L. L. wrote the original manuscript. All authors contributed to the discussions, reviewing, and final editing of the manuscript.

**Competing interests:** The authors declare that they have no competing interests.

**Data and materials availability:** All data are available in the main text or the supplementary materials.